\documentclass[11pt]{article}
\usepackage{latexsym}
\usepackage{amssymb}

\newtheorem{lemma}{Lemma}
\newtheorem{theorem}{Theorem}
\newtheorem{corollary}{Corollary}

\def\qed{\ \hfill$\Box$}
\def\mboxit#1{\mbox{{\it #1}}}

\title{{\bf Precoloring co-Meyniel graphs}}

\author{Vincent Jost\thanks{Laboratoire Leibniz-IMAG, 46 avenue
F\'elix Viallet, 38031 Grenoble Cedex, France $\{$vincent.jost,
benjamin.leveque, frederic.maffray$\}$@imag.fr}
\and%
Benjamin L\'ev\^eque$^\ast$
\and%
Fr\'ed\'eric Maffray$^\ast$\thanks{C.N.R.S.}}

\begin{document}

\maketitle

\begin{abstract}
The pre-coloring extension problem consists, given a graph $G$ and a
subset of nodes to which some colors are already assigned, in finding
a coloring of $G$ with the minimum number of colors which respects the
pre-coloring assignment.  This can be reduced to the usual coloring
problem on a certain contracted graph.  We prove that pre-coloring
extension is polynomial for complements of Meyniel graphs.  We answer
a question of Hujter and Tuza by showing that ``PrExt perfect'' graphs
are exactly the co-Meyniel graphs, which also generalizes results of
Hujter and Tuza and of Hertz.  Moreover we show that, given a
co-Meyniel graph, the corresponding contracted graph belongs to a
restricted class of perfect graphs (``co-Artemis'' graphs, which are
``co-perfectly contractile'' graphs), whose perfectness is easier to
establish than the strong perfect graph theorem.  However, the
polynomiality of our algorithm still depends on the ellipsoid method
for coloring perfect graphs.
\end{abstract}

\section{Introduction}

Often in applied optimization, one faces difficulty in the modelling
process because of the need to express some constraints that are not
extensively studied theoretically.  One type of such constraints is
that the organization of the system is partially fixed a priori, for
technical, historical or social reasons.  In terms of mathematical
programming, this can be interpreted as fixing the value of some
decision variables before the optimization process.  Although these
prerequirements cause the size of the problem to drop, they may
alterate the structural properties of the problem in such a way that
its complexity increases from polynomial to NP-hard.  The Pre-Coloring
extension problem, also called {\em PrExt}, is a good illustration of
this phenomenon.

For an integer $k$, a \emph{$k$-coloring} of the vertices of a graph
$G$ is the assignment of one element of $\{1, 2, \ldots, k\}$ (a
color) to each vertex of $G$ so that any two adjacent vertices receive
different colors.  Since each color class induces a stable set of $G$,
a coloring can also be seen as a partition of $V(G)$ into stable sets.
The smallest $k$ such that $G$ admits a $k$-coloring is the
\emph{chromatic number} of $G$, denoted by $\chi(G)$.  A
\emph{pre-coloring} of $G$ is a coloring of the vertices of an induced
subgraph of $G$, that is, a collection ${\cal Q}=\{C_1, \ldots, C_m\}$
of pairwise disjoint stable sets of $G$.  We say that a $k$-coloring
$\{S_1, \ldots, S_k\}$ of $G$ \emph{extends} ${\cal Q}$ if for each
$j=1, \ldots, m$ we have $C_j\subseteq S_j$.  The problem PrExt can be
defined as follows:

\begin{quote}
{\em Input:} A graph $G$, an integer $k$, and a pre-coloring ${\cal
Q}$ of $G$ using only colors from $\{1, \ldots, k\}$.

{\em Question:} Is there a $k$-coloring of $G$ that extends ${\cal Q}$
?
\end{quote}

PrExt is a generalization of coloring (which consists in taking ${\cal
Q}= \emptyset$), and it is more difficult than coloring since PrExt is
NP-complete even when restricted to bipartite graphs~\cite{Bod94,
Hujter93}, to interval graphs~\cite{Biro92} or to permutation
graphs~\cite{Jansen97}.  On the other hand, polynomial cases of PrExt
have been found, using several approaches, surveyed in~\cite{Tuza97}.

Given a pre-coloring ${\cal Q}=\{C_1, \ldots, C_m\}$ of $G$, we can
define a graph $G/{\cal Q}$ as follows.  Contract each $C_j$ into one
vertex $c_j$, with edges between $c_j$ and every vertex of $V(G)
\setminus (C_1\cup\cdots\cup C_m)$ that has at least one neighbour in
$C_j$.  Also add an edge between any two $c_j$'s.  The following
lemma, whose proof is obvious, shows how PrExt reduces to coloring via
contraction:

\begin{lemma}\label{modelisation}
For any graph $G$, and any pre-coloring ${\cal Q}$ of $G$, the set of
pre-coloring extensions of ${\cal Q}$ is in one-to-one correspondence
with the set of colorings of $G/{\cal Q}$.  In particular, the minimum
number of colors needed to extend ${\cal Q}$ is equal to $\chi(G/{\cal
Q})$.  \qed
\end{lemma}

Lemma~\ref{modelisation} shows that if we are able to solve the
coloring problem for $G/{\cal Q}$ then we are also able to solve PrExt
for $G$.  This paper attempts to give some insight on how this
translation works in general and to show in particular how it applies
to perfect graphs.  The latter point is summarized in the following
two theorems, which are the central results of this paper.  Before
presenting them we need some more definitions.  A \emph{cycle} of
length $p$ in a graph $G$ is a sequence of $p$ distinct vertices $v_1,
\ldots, v_p$ of $G$ such that $v_iv_{i+1}$ is an edge for each $i$
modulo $p$.  A \emph{chord} of the cycle is an edge $v_iv_j$ such that
$|j-i|\ge 2\bmod p$.  If the cycle has length at least $4$ and is
chordless, then it is also called a \emph{hole}.  If the cycle has
length at least five and has only one chord $p_1p_3$ (up to shifting
indices), then it is called a \emph{house}.  An \emph{antihole} is the
complementary graph of a hole.  A \emph{Meyniel graph}~\cite{markar,
mey} is a graph in which every odd cycle has at least two chords.  It
is easy to see that a graph $G$ is a Meyniel graph if and only if it
contains no odd hole and no house.

A graph $G$ is \emph{perfect} if, for every induced subgraph $H$ of
$G$, the chromatic number of $H$ is equal to the maximum clique size
in $H$.  Perfect graphs were introduced in 1960 by Berge,
see~\cite{ber85,ramree01}, who also conjectured that a graph is
perfect if and only if it does not contain an odd hole or an odd
antihole of length at least five.  This long-standing conjecture was
proved by Chudnovsky, Robertson, Seymour and Thomas~\cite{CRST}.  It
was known since \cite{markar, mey} that Meyniel graphs are perfect.

A \emph{prism} is a graph formed by three vertex-disjoint chordless
paths $P_1= u_0$-$\cdots$-$u_r$, $P_2= v_0$-$\cdots$-$v_s$, $P_3=
w_0$-$\cdots$-$w_t$ with $r,s,t\geq 1$, such that the sets $A=\{u_0,
v_0, w_0\}$ and $B=\{u_r, v_s, w_t\}$ are cliques and there is no edge
between the $P_i$'s other than the edges in $A$ and $B$.  Note that a
prism with $r=s=t=1$ is an antihole on six vertices.  A graph is an
\emph{Artemis} graph \cite{epsbook} if it contains no odd hole, no
antihole on at least five vertices, and no prism.  Artemis graphs are
perfect by the Strong Perfect Graph Theorem~\cite{CRST} but also by a
simpler result~\cite{Maffray02}.

Given a graph $G$, a \emph{co-coloring} of $G$ is a partition of
$V(G)$ into cliques of $G$.  For our purpose it will be more
convenient to talk about co-colorings than colorings.  Obviously
a co-coloring of $G$ is a coloring of $\overline{G}$, and all the
statements in this paper can be translated back and forth from
co-colorings to colorings by taking complementary graphs and
complementary classes of graphs.

Let ${\cal Q}=\{C_1, \ldots, C_m\}$ be a collection of pairwise
disjoint cliques of $G$.  Note that ${\cal Q}$ is a pre-coloring of
$\overline{G}$; so ${\cal Q}$ will be called a \emph{pre-co-coloring}
of $G$.  We denote by $G^{\cal Q}$ the graph obtained by the operation
of \emph{co-contraction} defined as follows.  Each element $C_j$ of
${\cal Q}$ is contracted into one vertex $c_j$.  A vertex of
$G\setminus (C_1\cup\cdots\cup C_m)$ is adjacent to $c_j$ in $G^{\cal
Q}$ if and only if it is adjacent in $G$ to every vertex of $C_j$; and
there is no edge between any two $c_j$'s.  Clearly, $G^{\cal Q}$ is
the complementary graph of $\overline{G}/{\cal Q}$.

\newpage

\begin{theorem}\label{thm1}
The co-contraction $G^{\cal Q}$ of $G$ is a perfect graph for every
pre-co-coloring ${\cal Q}$ if and only if $G$ is a Meyniel graph.
\end{theorem}
\begin{theorem}\label{thm2}
If $G$ is a Meyniel graph and ${\cal Q}$ is any pre-co-coloring of
$G$, then the co-contracted graph $G^{\cal Q}$ is an Artemis graph.
\end{theorem}
Theorem~\ref{thm1} has a nice algorithmic consequence:
\begin{corollary}\label{corcom}
PrExt is polynomial for co-Meyniel graphs.
\end{corollary}
{\it Proof.} PrExt on co-Meyniel graphs is equivalent to co-PrExt on
Meyniel graphs.  Given a Meyniel graph $G$ and a co-coloring ${\cal
Q}$ of $G$, the co-contracted graph $G^{\cal Q}$ is perfect by
Theorem~\ref{thm1}.  One can use a polynomial-time algorithm
~\cite{Grot, Schrijver} to find an optimum co-coloring for $G^{\cal
Q}$.  From this co-coloring, using Lemma~\ref{modelisation}, one
deduces an optimal pre-co-coloring extension of ${\cal Q}$ for $G$.
\qed

Corollary~\ref{corcom} contains and unifies several previously known
cases of polynomiality of PrExt~\cite{Hujter93, Hujter96, Tuza97}:
split graphs, cographs, P5-free bipartite graphs, complements of
bipartite graphs, and the case of a co-Meyniel graph where every
pre-color class has size $1$~\cite{hertz}.  The proof of
corollary~\ref{corcom} can also be used to derive a more general
algorithmic consequence of Lemma~\ref{modelisation}.  Given a class of
graphs ${\cal G}$, let its ``co-contraction closure'' ${\cal G}^+$ be
the class of all graphs obtained by the co-contraction of any
pre-co-colored graph in ${\cal G}$.
\begin{corollary}\label{cor:complexity}
Let ${\cal G}$ be a class of graph.  If co-coloring is polynomial on
graph class ${\cal G}^+$, then co-PrExt is polynomial on ${\cal G}$.
\end{corollary}

One use of Corollary~\ref{cor:complexity} is to reduce co-PrExt on a
given class of graph ${\cal G}$ to asking ``what is ${\cal G}^+$'' and
then solving co-coloring on ${\cal G}^+$.  Unfortunately this strategy
may fail in general, because even if we are able to describe ${\cal
G}^+$, we do not necessarily know the complexity of co-coloring on
${\cal G}^+$.  It may be more fruitful to try to translate results
from co-coloring to co-PrExt.  For instance, in this paper, we want to
apply the ellipsoid method, which allows to (co)-color perfect graphs
in polynomial time~\cite{Grot, Schrijver}.  So we can ask: ``what is
the class ${\cal G}$ such that ${\cal G}^+ = \mboxit{Perfect}$ ?''.
Unfortunately, such a class does not exist.  To clarify this point,
let us give another definition. Given a class ${\cal G}$, let ${\cal
G}^-$ be the set of graphs $G$ such that $G^{\cal Q}$ belongs to
${\cal G}$ for every ${\cal Q}$.
It is easy to see that every class ${\cal G}$ of graphs satisfies
$({\cal G}^-)^+ \subseteq {\cal G} \subseteq ({\cal G}^+)^-$.
Theorem~\ref{thm1} says that $\mboxit{Perfect}^- = \mboxit{Meyniel}$,
and Theorem~\ref{thm2} says that $\mboxit{Meyniel}^+ \subseteq
\mboxit{Artemis}$, which is a strict subclass of perfect graphs.  It
follows that $(\mboxit{Perfect}^-)^+ \neq \mboxit{Perfect}$, and
consequently that there is no class ${\cal G}$ of graphs such that
${\cal G}^+ = \mboxit{Perfect}$.

This discussion suggests a weaker but more directly usable version of
corollary~\ref{cor:complexity}.
\begin{corollary}\label{cor:complexity2}
Let ${\cal G}$ be a class of graph.  If co-coloring is polynomial on
graph class ${\cal G}$, then co-PrExt is polynomial on ${\cal G}^-$.
\end{corollary}

Let us also note that Theorem~\ref{thm1} answers the question of
Hujter and Tuza~\cite{Hujter96} to characterize ``PrExt-Perfect
graphs'' which, in our language, was precisely to characterize the
class $\mboxit{Perfect}^-$.  Indeed, Hujter and Tuza's so-called
``core condition'' turns out to be equivalent to the clique condition
$\chi(G/{\cal Q}) \geq \omega(G/{\cal Q})$.  They called a graph $G$
``PrExt-perfect'' if both $G/{\cal Q}$ is perfect for every ${\cal Q}$
and the core condition is sufficient for extendibility.  A consequence
of our Lemma~$1$ is that this second condition is redundant, because
perfection implies sufficiency of the clique condition in $G/{\cal
Q}$.  Hence their PrExt-perfect graphs coincide with
$\mboxit{Perfect}^-$.

%
%

\section{Proof of Theorems~\ref{thm1} and~\ref{thm2}}\label{sec:proofs}

Throughout this section, $G$ is a graph and ${\cal Q}$ is a
pre-co-coloring of $G$.

One way in Theorem~\ref{thm1} (namely, $\mboxit{Perfect}^- \subseteq
\mboxit{Meyniel}$) is easy:

\begin{lemma}\label{necessity}
If the co-contracted graph $G^{\cal Q}$ is perfect for all
pre-co-coloring ${\cal Q}$ of $G$, then $G$ is Meyniel.
\end{lemma}
{\it Proof.} If $G$ contains an odd hole, then the graph $G^\emptyset$
contains this hole.  If $G$ contains a house with the chord $xy$ and
$G$ contains no odd hole, then the house has odd length and the
co-contracted graph $G^{\{\{x\}, \{y\}\}}$ contains an odd hole.  So
$G$ must be a Meyniel graph for $G^{\cal Q}$ to be perfect for every
${\cal Q}$.  \qed

The rest of this section is devoted to the study of
$\mboxit{Meyniel}^+$.

\begin{lemma}\label{meyniel}
In a Meyniel graph $G$,  let $P = p_0$-$\cdots$-$p_n$ be a chordless
path and $x$ be a vertex of $V(G)\setminus V(P)$ which sees $p_0$ and
$p_n$.  Then either $x$ sees every vertex of $P$,  or $n$ is even and
$N(x)\cap V(P)\subseteq \{p_{2i} \mid i=0,  \ldots,  n/2\}$.
\end{lemma}
{\it Proof.} Call segment any subpath of length at least $1$ of $P$
whose two endvertices see $x$ and whose interior vertices do not.
Since $p_0$ and $p_n$ see $x$, path $P$ is partitioned into its
segments.  Let $p_h$-$\cdots$-$p_j$ be any segment with $j-h\ge 2$.
Then $x, p_h, \ldots, p_j$ induce a hole, so $j-h$ is even.  Thus
every segment has length either even or equal to $1$.  Suppose that
there is a segment of length $1$ and a segment of even length.  Then,
there are consecutive such segments, that is, up to symmetry, there
are integers $0<h<j\le n$ such that $p_{h-1}$-$p_h$ is a segment of
length $1$ and $p_h$-$\cdots$-$p_j$ is a segment of even length; but
then $x, p_{h-1}, p_h, \ldots, p_j$ induce a house, a contradiction.
Thus either all segments have length $1$ (i.e., $x$ sees every vertex
of $P$), or they all have even length, and the lemma holds.  \qed

\begin{lemma}\label{consecutive}
In a Meyniel graph $G$,  let $H$ be an even hole and $x$ be a vertex of
$V(G)\setminus V(H)$ that sees two consecutive vertices of $H$.  Then
$x$ sees either all vertices of $H$ or exactly three consecutive
vertices of $H$.
\end{lemma}
{\it Proof.} Let $x_1, \ldots,  x_n$ be the vertices of $H$ ordered
cyclically.  Suppose that $x$ sees $x_1$ and $x_2$ but not all
vertices of $H$,  and let $x_i$ be a vertex of $H$ that is not seen by
$x$.  Suppose that $x$ sees a vertex $x_j$ with $j\notin \{1,  2,  3,
n\}$.  By symmetry we can assume that $i<j$.  Then either
$x_1$-$\cdots$-$x_j$ or $x_2$-$\cdots$-$x_j$ is an odd chordless path,
and in either case $x$ sees the two endvertices and not all vertices
of that path,  a contradiction to Lemma~\ref{meyniel}.  So $N(x)\cap
V(H) \subseteq \{x_1,  x_2,  x_3,  x_n\}$.  If $x$ sees none of $x_3,
x_n$,  then $V(H)\cup\{x\}$ induces a house,  a contradiction.  If $x$
sees both $x_3,  x_n$,  then $x,  x_3, \ldots,  x_n$ induce an odd hole,  a
contradiction.  So $x$ sees exactly one of $x_3,  x_n$,  and the lemma
holds.  \qed

\begin{lemma}\label{pqz}
In a Meyniel graph $G$,  let $Q$ be a clique,  $P = p_0$-$\cdots$-$p_n$
be a chordless path in $G\setminus Q$,  and $z$ be a vertex not in
$Q\cup V(P)$.  Suppose that $z$ and $p_0$ see all vertices of $Q$,
that $z$ does not see $p_1$,  and that some vertex $q\in Q$ sees $p_0$
and $p_n$ and not $p_1$.  Then $p_n$ sees all vertices of $Q$.
\end{lemma}
{\it Proof.} We prove this lemma by induction on $n$.  Suppose that
some vertex $q'\in Q$ does not see $p_n$.  By Lemma~\ref{meyniel}
applied to $P$ and $q$,  since $q$ sees $p_0,  p_n$ and not $p_1$,  path
$P$ has even length and $N(q)\cap V(P) \subseteq \{p_{2i} \mid i=1,
\ldots,  n/2\}$.  Let $j$ be the largest integer such that $q$ sees
$p_j$ and $j<n$.  Since $j$ is even,  $p_j,  \ldots,  p_n,  q$ induce an
even hole $C$.  Suppose that $j=0$.  Then $q'$ sees $q,  p_0$ on $C$
and not $p_n$,  so Lemma~\ref{consecutive} implies that $q'$ sees $p_1$
and none of $p_2,  \ldots,  p_n$.  Call $C'$ the even hole induced by
$(V(C)\setminus p_0) \cup \{q'\}$.  Vertex $z$ sees $q,  q'$ of $C'$
and not $p_1$,  so Lemma~\ref{consecutive} implies that $z$ sees $p_n$
and none of $p_1,  \ldots,  p_{n-1}$.  But then $V(C)\cup \{z\}$ induces
a house,  a contradiction.  So $j\ge 2$.  By the induction hypothesis,
$p_j$ sees all vertices of $Q$.  Then $q'$ sees $q,  p_j$ on $C$ and
not $p_n$,  so Lemma~\ref{consecutive} implies that $q'$ sees $p_{j+1}$
and none of $p_{j+2},  \ldots,  p_n$.  But then $p_0,  q,  q',  p_{j+1},
\ldots,  p_n$ induce a house,  a contradiction.  \qed

\begin{lemma}\label{notadj}
In a Meyniel graph $G$,  let $Q$ be a clique,  $X$ be a connected set of
vertices of $G\setminus Q$,  and $z$ be a vertex not in $Q\cup X$.
Suppose that $z$ sees all the vertices of $Q$ and none of $X$,  and
that each vertex of $X$ has a non-neighbour in $Q$.  Then some vertex
of $Q$ has no neighbour in $X$.
\end{lemma}
{\it Proof.} We prove the lemma by induction on the size of $X$.  If
$|X|=1$ there is nothing to prove,  so assume $|X|\ge 2$.  Let $x, x'$
be two distinct vertices of $X$ such that $X\setminus \{x\}$ and
$X\setminus \{x'\}$ are connected (for example let $x, x'$ be two
leaves of a spanning tree of $X$).  By the induction hypothesis,  there
are vertices $q, q'$ of $Q$ such that $q$ has no neighbour in
$X\setminus \{x\}$ and $q'$ has no neighbour in $X\setminus \{x'\}$.
If either $q$ does not see $x$ or $q'$ does not see $x'$,  then the
lemma holds,  so suppose that $q$ sees $x$ and $q'$ sees $x'$.  Let $P$
be a shortest path from $x$ to $x'$ in $X$.  Then either $V(P)\cup
\{q, q'\}$ induces an odd hole or $V(P)\cup \{q, q', z\}$ induces a
house,  a contradiction.  \qed

\begin{lemma}\label{antihole}
Let $G$ be a Meyniel graph and ${\cal Q}$ be a precocoloring of $G$.
Then the cocontracted graph $G^{\cal Q}$ contains no antihole of size
at least $6$.
\end{lemma}
{\it Proof.} Suppose that $G^{\cal Q}$ contains an antihole $A$ of
size at least $6$.  Since the cliques of ${\cal Q}$ are cocontracted
into a stable set,  there are at most two vertices in $A$ that result
from the cocontraction of a clique and if there are two such vertices
they are consecutive in the cyclic ordering of $\overline{A}$.  If
there are five consecutive vertices of $A$ that do not result from the
cocontraction of a clique,  then these five vertices form a house of
$G$,  a contradiction.  So there are no such five vertices,  which
implies that $A$ is of size six and has exactly two cocontracted
vertices.  Let $x_1,  \ldots,  x_6$ be the vertices of $A$ ordered
cyclically,  such that $x_1,  x_2$ are the cocontracted vertices.  Let
$C_1$ be the clique whose cocontraction results in $x_1$.  Since $x_1$
and $x_6$ are not adjacent,  $x_6$ does not see all the vertices of
$C_1$,  so there is a vertex $q_1$ of $C_1$ that does not see $x_6$ in
$G$.  Then $q_1,  x_3,  x_4,  x_5,  x_6$ induce a house in $G$,  a
contradiction.  \qed

\begin{lemma}\label{oddhole}
Let $G$ be a Meyniel graph and ${\cal Q}$ be a precocoloring of $G$.
Then the cocontracted graph $G^{\cal Q}$ contains no odd hole.
\end{lemma}
{\it Proof.} We prove the lemma by induction on $m=|{\cal Q}|$.  If
$m=0$,  then $G^{\cal Q}=G$ and the lemma holds.  So assume $m>0$ and
let ${\cal Q}=\{C_1,  \ldots,  C_m\}$.  Suppose that $G^{\cal Q}$
contains an odd hole ${\cal H}$.  Let $x_1,  \ldots,  x_n$ be the
vertices of ${\cal H}$ ordered cyclically.  For each $j=1,  \ldots,  m$,
we may assume that the vertex that results from the cocontraction of
$C_j$ lies in ${\cal H}$,  for otherwise ${\cal H}$ is an odd hole in
$G^{\cal Q}\setminus \{C_j\}$,  which contradicts the induction
hypothesis.  So let us call $x_{i_j}$ the vertex of ${\cal H}$ that
results from the cocontraction of $C_j$,  and assume without loss of
generality that $1 < i_1<i_2<\cdots <i_m\le n$.

Suppose that $m=1$.  We may assume that $i_1=1$.  Since $x_3$ and
$x_1$ are not adjacent in $G^{\cal Q}$,  $x_3$ does not see all
vertices of $C_1$ in $G$,  so there is a vertex $q_1$ of $C_1$ that
does not see $x_3$ in $G$.  Then the path $P=x_2$-$\cdots$-$x_n$ is
chordless and odd,  and $q_1$ sees both endvertices of $P$ and misses
vertex $x_3$ of $P$,  a contradiction to Lemma~\ref{meyniel}.
Therefore $m\ge 2$.

The cocontracted vertices $x_{i_1},  \ldots ,  x_{i_m}$ form a stable
set in ${\cal H}$.  So for all $j$,  we have $i_{j+1}-i_j \geq 2$.
Since $n$ is odd and $m\ge 2$,  there exists $j$ such that $i_{j+1}-
i_j$ is odd (and so $i_{j+1}-i_j\ge 3$).  We can assume without loss
of generality that $i_2-i_1$ is odd and $i_1=1$ (so $i_2\ge 4$).  Let
$R$ be the odd path $x_2$-$\cdots$-$x_{i_2-1}$.

Since $x_3$ and $x_1$ are not adjacent in $G^{\cal Q}$,  there is a
vertex $q_1$ of $C_1$ that does not see $x_3$ in $G$.  Likewise,  there
is a vertex $q_2$ of $C_2$ that does not see $x_{i_2-2}$ in $G$.
Moreover,  if $m\ge 3$,  we can apply Lemma~\ref{notadj} to the clique
$C_j$,  the connected set $R$ and $x_{i_j-1}$,  which implies that:
\begin{equation}\label{cl3}
\mboxit{\it For $j=3,  \ldots,  m$,  there is a vertex $q_j\in C_j$ that
sees no vertex of $R$.}
\end{equation}

Now we select vertices $y_1,  \ldots,  y_n$ of $G$ as follows.  For
$k=1,  \ldots,  n$,  if there exists $j$ such that $i_j=k$,  let $y_k=
q_j$; else let $y_k= x_k$.  The selected vertices $y_1,  \ldots,  y_n$
form an odd cycle $H$ of $G$.  Note that,  in $H$,  vertex $y_2$ is
adjacent only to $y_1,  y_3$ and possibly to $y_{i_2}=q_2$,  by
(\ref{cl3}).
\begin{equation}\label{clu}
\mboxit{\it Every neighbour of $q_1$ in $V(H) \setminus \{y_2,  y_n\}$
is in $\{q_2,  \ldots,  q_m\}$.}
\end{equation}
For let $u$ be a neighbour of $q_1$ in $V(H) \setminus \{y_2,  y_n\}$.
Since $x_2$ and $x_{i_2}$ are not adjacent in $G^{\cal Q}$,  there is a
vertex $q'_2$ of $C_2$ that does not see $y_2$ in $G$.  The subgraph
of $G$ induced by $V(H)\cup\{q'_2\}\setminus\{y_1,  y_n,  q_2\}$ is
connected,  so it contains a shortest path $U$ from $y_2$ to $u$.
Since $y_3$ is the only neighbour of $y_2$ in that subgraph,  $y_3$
lies on $U$ as the neighbour of $y_2$.  Now Lemma~\ref{pqz} can be
applied to $C_1$,  $U$ and $y_n$,  which implies that $u$ sees all of
$C_1$.  Then $u$ must be in $\{q_2,  \ldots,  q_m\}$ for otherwise $q_1$
and $u$ would be adjacent in $G^{\cal Q}$.  So (\ref{clu}) holds.
Likewise:
\begin{equation}\label{clv}
\mboxit{\it Every neighbour of $q_2$ in $V(H) \setminus \{y_{i_2-1},
y_{i_2+1}\}$ is in $\{q_1,  q_3,  \ldots,  q_m\}$.}
\end{equation}
Now each of $y_2,  \ldots,  y_{i_2-1}$ has degree $2$ in $H$.

Let us color blue some vertices of the path ${\cal H}\setminus x_1$
$=x_2$-$\cdots$-$x_n$ of $G^{\cal Q}$ as follows.  Vertices $x_2$ and
$x_n$ are colored blue.  For $j=2,  \ldots,  m$,  vertex $x_{i_j}$ is
colored blue if and only if all vertices of the corresponding clique
$C_j$ see $q_1$.  All other vertices of ${\cal H}$ are uncolored.
Call blue segment any subpath of length at least $1$ of ${\cal H}
\setminus x_1$ whose two endvertices are blue and whose interior
vertices are uncolored.  Since ${\cal H} \setminus x_1$ has odd length
and its endvertices are blue,  it has an odd blue segment.  Let
$x_h$-$\cdots$-$x_i$ be any odd blue segment,  with $2\le h<i\le n$.
Suppose that $i-h\ge 3$.  Then (\ref{clu}) implies that $q_1,  x_h,
x_{h+1},  \ldots,  x_{i-1},  x_i$ induce an odd hole in $G^{\cal
Q}\setminus \{C_1\}$,  which contradicts the induction hypothesis on
$|{\cal Q}|$.  So we must have $i-h=1$.  Since $i_{j+1}-i_j\ge 2$ for
all $j$ and $i_2\ge 4$,  this is possible only if $h=n-1$.  This
implies that $x_{n-1}$-$x_n$ is the only odd blue segment,  and that
every blue vertex $x_k$ different from $x_n$ has even $k$.

Likewise,  we color red some vertices of the path ${\cal H}\setminus
x_{i_2}$ of $G^{\cal Q}$ as follows.  Vertices $x_{i_2-1}$ and
$x_{i_2+1}$ are colored red.  For $j=1,  2,  \ldots,  m$ and $j\neq 2$,
vertex $x_{i_j}$ is colored red if and only if all vertices of the
corresponding clique $C_j$ see $q_2$.  Call red segment any subpath of
length at least $1$ of ${\cal H} \setminus x_{i_2}$ whose two
endvertices are red and whose interior vertices are not red.  Just
like in the preceding paragraph,  we obtain that
$x_{i_2+1}$-$x_{i_2+2}$ is the only odd red segment,  and that every
red vertex $x_l$ different from $x_{i_2-1}$ and $x_{i_2+1}$ has either
even $l$ or $l=1$.

If $i_2=n-1$,  then $m=2$ and $V(R)\cup\{q_1,  q_2,  x_n\}$ induces an
odd hole (if $q_1,  q_2$ are not adjacent) or a house (if $q_1,  q_2$
are adjacent) in $G$,  a contradiction.  So suppose $i_2\le n-3$.
Since $x_{i_2+2}$ is red and $x_{n-1}$ is blue,  there is a subpath
$x_k$-$\cdots$-$x_l$ of $x_{i_2+2}$-$\cdots$-$x_{n-1}$ such that $x_k$
is red,  $x_l$ is blue,  and no interior vertex of $x_k$-$\cdots$-$x_l$
is colored.  By the preceding paragraphs,  both $k, l$ are even.  If
$k=l$,  then (\ref{clu}) implies that there is a clique $C_j$ such that
$k=i_j$,  and then $V(R)\cup\{q_1,  q_2,  q_j\}$ induces an odd hole (if
$q_1,  q_2$ are not adjacent) or a house (if $q_1,  q_2$ are adjacent)
in $G$,  a contradiction.  So $k\neq l$.  If $q_1,  q_2$ are adjacent,
then (\ref{clu}) and (\ref{clv}) imply that $\{q_1,  x_k,  \ldots,  x_l,
q_2\}$ induces an odd hole in $G^{\cal Q} \setminus \{C_1,  C_2\}$,  a
contradiction to the induction hypothesis on $|{\cal Q}|$.  If $q_1,
q_2$ are not adjacent then $V(R)\cup\{q_1,  x_k,  \ldots,  x_l,  q_2\}$
induces an odd hole in $G^{\cal Q} \setminus \{C_1,  C_2\}$,  again a
contradiction.  This completes the proof of the lemma.  \qed


\begin{lemma}\label{prism}
Let $G$ be a Meyniel graph and ${\cal Q}$ be a precocoloring of $G$.
Then the cocontracted graph $G^{\cal Q}$ contains no prism.
\end{lemma}
{\it Proof.} Suppose that $G^{\cal Q}$ contains a prism $K$ formed by
paths $P_1= u_0$-$\cdots$-$u_r$, $P_2= v_0$-$\cdots$-$v_s$, $P_3=
w_0$-$\cdots$-$w_t$, with $r,s,t\ge 1$, and with triangles $A=\{u_0,
v_0, w_0\}$ and $B=\{u_r, v_s, w_t\}$.  By Lemma~\ref{antihole}, $K$
is not an antihole on $6$ vertices, so we can assume that one of
$r,s,t$ is not equal to $1$.  Let ${\cal Q}=\{C_1, \ldots, C_m\}$.  We
have $m\ge 1$ since a Meyniel graph contains no prism, because a prism
contains a house.  For each $j=1, \ldots, m$, call $c_j$ the vertex of
$G^{\cal Q}$ that results from the cocontraction of $C_j$, and let
$C=\{c_1, \ldots, c_m\}$.  By Lemma~\ref{oddhole}, $G^{\cal Q}$
contains no odd hole, thus $r, s, t$ have the same parity.  Note that
$V(K)\setminus C\subset V(G)$, and, since $C$ is a stable set,
$N(c_j)\subset V(G)$ for each $j=1, \ldots, m$.  We claim that:
\begin{equation}\label{cla1}
\mboxit{\it $|A\cap C|= 1$ and $|B\cap C|=1$.}
\end{equation}
Note that $|A\cap C|\le 1$ and $|B\cap C|\le 1$ since $A, B$ are
cliques and $C$ is a stable set of $G^{\cal Q}$.  Now, suppose up to
symmetry that $A\cap C=\emptyset$.  For each $j=1, \ldots, m$, we can
apply Lemma~\ref{notadj} in $G$ to the clique $C_j$, the connected set
$A\setminus N(c_j)$, and some neighbour $z$ of $c_j$ in $K$ (more
precisely: if $c_j=u_i$ with $i<r$ then take $z=c_{i+1}$; if $c_j=u_r$
and either $s\ge 2$ or $t\ge 2$, take $z=v_s$ or $z=w_t$ respectively;
if $c_j=u_r$ and $s=t=1$, then $r\ge 3$ and take $z=u_{r-1}$; a
similar such $z$ exists if $c_j\in V(P_2) \cup V(P_3)$).
Lemma~\ref{notadj} implies that there is a vertex $q_j\in C_j$ that
sees no vertex of ${\cal A}\setminus N(c_j)$.  Let $P$ be the subgraph
of $G$ induced by $(V(K)\setminus C)\cup \{q_1, \ldots, q_m\}$.  Let
$u'_1$ be the neighbour of $u_0$ in $P\setminus \{v_0, w_0\}$ (so
$u'_1$ is either $u_1$ or some $q_j$), and similarly let $v'_1$ be the
neighbour of $v_0$ in $P\setminus \{u_0, w_0\}$.  Let $R$ be a
shortest path from $u'_1$ to $v'_1$ in $P \setminus \{u_0, v_0,
w_0\}$.  Then $V(R)\cup \{u_0, v_0, w_0\}$ induces a house in $G$, a
contradiction.  So (\ref{cla1}) holds.

By (\ref{cla1}) and up to symmetry we may assume that $u_0=c_1$,  and
so $v_0,  w_0$ are vertices of $G$.  As above,  by Lemma~\ref{notadj},
for $j=2,  \ldots,  m$,  we can select a vertex $q_j$ in $C_j$ that
misses all of $\{v_0,  w_0\}\setminus N(c_j)$.  We claim that:
\begin{equation}\label{cla2}
\mboxit{\it Vertices $v_1$ and $w_1$ of $G^{\cal Q}$ are in $C$.}
\end{equation}
For suppose,  up to symmetry,  that $v_1$ is not in $C$.  Then we can
select a vertex $q'_1\in C_1$ that misses $v_1$.  Let $P$ be the
subgraph of $G$ induced by $(V(K)\setminus C)\cup \{q'_1,  q_2,  \ldots,
q_m\}$.  Let $R$ be a shortest path from $q'_1$ to $v_1$ in
$P\setminus \{v_0,  w_0\}$.  Then $R$ has length at least $2$ and
$V(R)\cup \{v_0,  w_0\}$ induces a house in $G$,  a contradiction.  So
(\ref{cla2}) holds.

By (\ref{cla2}),  we may assume that $v_1=c_2$ and $w_1=c_3$.  Recall
that $q_2$ is a vertex of $C_2$ that misses $w_0$,  and $q_3$ is a
vertex of $C_3$ that misses $v_0$.  Since $v_1$ and $w_1$ are not
adjacent,  the lengths $s, t$ of $P_2, P_3$ cannot both be equal to $1$;
thus let us assume up to symmetry that $s\ge 2$.  We claim that:
\begin{equation}\label{cla3}
\mboxit{\it Vertex $q_2$ is adjacent to all of $C_1$.}
\end{equation}
For suppose that $q_2$ is not adjacent to some vertex $q''_1\in C_1$.
Let $P$ be the subgraph of $G$ induced by $(V(K)\setminus C)\cup
\{q''_1,  q_2,  \ldots,  q_m\}$.  Let $R$ be a shortest path from $q''_1$
to $q_2$ in $P \setminus \{v_0,  w_0\}$.  Then $R$ has length at least
$2$ and $V(R)\cup \{v_0,  w_0\}$ induces a house in $G$,  a
contradiction.  So (\ref{cla3}) holds.

Now let $q_1$ be any vertex of $C_1$,  and let $P$ be the subgraph of
$G$ induced by $(V(K)\setminus C)\cup \{q_1,  \ldots,  q_m\}$.  We claim
that:
\begin{equation}\label{cla4}
\mboxit{\it Every neighbour of $q_2$ in $V(P)\setminus \{v_0,  v_2\}$
is in $\{q_1,  \ldots,  q_m\}$.}
\end{equation}
For let $x$ be a neighbour of $q_2$ in $V(P)\setminus \{v_0,  v_2\}$.
We can suppose that $x\neq q_1$.  The subgraph $P\setminus \{q_1,  q_2,
v_2\}$ is connected,  so it contains a shortest path $X$ from $v_0$ to
$x$.  Since $v_0$ has no neighbour in $V(P)\setminus\{q_1,  q_2,
w_0\}$,  $w_0$ lies on $X$ as the neighbour of $v_0$.  Now
Lemma~\ref{pqz} can be applied to $C_2$,  $X$ and $v_2$,  which implies
that $x$ sees all of $C_2$.  This means that $x$ is in $\{q_1,  \ldots,
q_m\}$ for otherwise $v_1$ and $x$ would be adjacent in $G^{\cal Q}$.
So (\ref{cla4}) holds.

In $G^{\cal Q}$,  let us mark some vertices of $K\setminus v_1$ as
follows.  Vertices $v_0$ and $v_2$ are marked.  For $j=1,  \ldots,  m$
and $j\neq 2$,  vertex $c_j$ is marked if and only if in $G$ vertex
$q_2$ sees all vertices of the corresponding clique $C_j$ in $G$.  All
other vertices of $K$ are unmarked.  Call segment any subpath of
length at least $1$ of $K \setminus v_1$ whose two endvertices are
marked and whose interior vertices are unmarked.  Suppose there exists
an odd segment $X$ of length $\ge 3$.  Then $V(X)\cup\{q_2\}$ induces
an odd hole in $G^{{\cal Q}\setminus \{C_2\}}$,  which contradicts
Lemma~\ref{oddhole}.  So every segment has length even or equal to
$1$.  Note that $V(P_1) \cup V(P_2) \setminus \{v_0,  v_1\}$ induces a
chordless path,  of odd length (because $r, s$ have the same parity),
and its two extremities are marked; so this path contains an odd
segment,  which as noted above has length $1$.  Call $y$ the neighbour
of $v_2$ on that path.  Note that we have either $s\ge 3$ and $y=v_3$
or $s=2$ and $y=u_r$.  By (\ref{cla4}) and the fact that vertices of
$C$ are pairwise non-adjacent,  the only possible segment of length one
is $v_2$-$y$,  so $y$ is marked,  and (\ref{cla4}) implies $y\in C$.
Suppose that $s\ge 3$.  Then $V(P_3) \cup V(P_2)\setminus \{v_1,
v_2\}$ induces a chordless odd path,  whose two extremities are marked,
so it contains a segment of length $1$.  The only possible such
segment is $v_0$-$w_0$,  so $w_0$ is marked,  and (\ref{cla4}) implies
$w_0\in C$,  which contradicts (\ref{cla1}).  So $s=2$ and $y=u_r$.
Now,  since $B$ contains $u_r$ it cannot contain another vertex of $C$,
so $w_t$ is not in $C$,  which by (\ref{cla2}) implies $t\ge 2$.  Now
symmetry between $s$ and $t$ is restored,  and as above we can prove
that $t=2$ and $q_3$ is adjacent to $u_r$.  But then $q_2,  v_0,  w_0,
q_3,  u_r$ induce an odd hole or a house in $G$,  a contradiction.  \qed


Now Lemmas~\ref{antihole}, \ref{oddhole} and~\ref{prism} imply that
$G^{\cal Q}$ is an Artemis graph, which proves Theorem~\ref{thm2}.
Theorem~\ref{thm2} and Lemma~\ref{necessity} imply Theorem~\ref{thm1}.

\section{Concluding remarks}\label{sec:epilogue}

This is still not the end of the story.  The general method is as
follows.  Assume that we want to apply a co-coloring algorithm $A$
whose validity is proved on a class ${\cal G}$.  Then we can use $A$
for the problem co-PrExt on ${\cal G}^-$.  Since we know that $({\cal
G}^-)^+ \subseteq {\cal G}$ only, we can wonder what is the class
$({\cal G}^-)^+$, because there might exist algorithms that are better
than $A$ to co-color graphs in $({\cal G}^-)^+$ (or for solving
co-PrExt on ${\cal G}^-$).  Here we proved that $\mboxit{Perfect}^- =
\mboxit{Meyniel}$ and that $\mboxit({Perfect}^-)^+ =
\mboxit{Meyniel}^+ \subseteq \mboxit{Artemis} \subsetneq
\mboxit{Perfect}$.  Improving from Perfect to Artemis (in the last
strict inclusion) has two interesting aspects: First, perfection of
Artemis graphs~\cite{Maffray02} is easier to establish than perfection
of Berge graphs.  Second, since co-PrExt is polynomial on Meyniel
graphs with the ellipsoid method, the question arises of finding a
combinatorial algorithm for this question.  However we do not have an
answer for this and we leave it as an open problem.


The scope of applications of Lemma~\ref{modelisation} might not be
completely exploited yet: for instance the computational complexity
equivalence may work in any computational class ${\cal C}$ (APX, NP,
\dots), provided that the reduction in the proof of
Lemma~\ref{modelisation} preserves the properties of ${\cal C}$.  For
instance, it is an $AP$-reduction (see~\cite{aus} for the background
concerning approximability, both in general and concerning coloring
problems); so approximability results can be transposed from coloring
on $\cal{G}$ to PrExt on ${\cal G}^-$.  An extension could be the
converse of Corollary~\ref{cor:complexity}; this would allow for a
translation of inaproximability results from coloring to precoloring
extension.  The difficulty here is, given a graph class $\cal{G}$ and
a graph $G$ (not necessarily in $\cal{G}^+$), to find a graph $ H \in
\cal{G}$ and a precoloring $\cal{Q}$ of $H$ such that $H / \cal{Q} =
G$ or to certify that there is no such pair $H, \cal{Q}$.  The
complexity of this problem is open, even if we restrict $\cal{G}$ to
be $\textit{Meyniel}$ (note here that $\textit{Meyniel}^+$ is not even
well characterized yet).

\end{document}